\documentclass[10pt,a4paper,english,twocolumn,
aps,
floatfix,superscriptaddress
]{revtex4}
\usepackage[T1]{fontenc}
\usepackage[latin9]{inputenc}
\usepackage{amsmath}
\usepackage{amssymb}
\usepackage{esint}
\usepackage{color}
\usepackage[colorlinks=true,citecolor=blue]{hyperref}

\renewcommand{\b}[1]{\mathbf{ #1}}									%vettori grasetti
	
\newcommand{\cop}[1]{{#1^{\dagger}}\!}  						%creation opeator			
\newcommand{\aop}[1]{{#1}}

\newcommand{\beq}{\begin{equation}}
\newcommand{\eeq}{\end{equation}}

\makeatletter

\@ifundefined{textcolor}{}
{%
 \definecolor{BLACK}{gray}{0}
 \definecolor{WHITE}{gray}{1}
 \definecolor{RED}{rgb}{1,0,0}
 \definecolor{GREEN}{rgb}{0,1,0}
 \definecolor{BLUE}{rgb}{0,0,1}
 \definecolor{CYAN}{cmyk}{1,0,0,0}
 \definecolor{MAGENTA}{cmyk}{0,1,0,0}
 \definecolor{YELLOW}{cmyk}{0,0,1,0}
}

\usepackage{graphicx}  % needed for figures
\usepackage{dcolumn}   % needed for some tables
\usepackage{bm}        % for math
\usepackage{babel}

\makeatother

\begin{document}

\title{Solvable two-dimensional superconductors with $l$-wave pairing}

\author{Luca Lepori}
\email[correspondence at: ]{llepori81@gmail.com}
\affiliation{Istituto Italiano di Tecnologia, Graphene Labs, Via Morego 30, I-16163 Genova,~Italy.}
\affiliation{Dipartimento di Scienze Fisiche e Chimiche, Universit\`a dell'Aquila, via Vetoio,
I-67010 Coppito-L'Aquila, Italy.}
\affiliation{INFN, Laboratori Nazionali del Gran Sasso, Via G. Acitelli, 22, I-67100 Assergi (AQ), Italy.}

\author{Marco Roncaglia}
\email[correspondence at: ]{marco.roncaglia.it@gmail.com}
\affiliation{Physics Department and Research Center OPTIMAS, University of Kaiserslautern, Erwin-Schroedinger-Str. 46, 
67663 Kaiserslautern, Germany.}

\begin{abstract}
We analyze a family of  two-dimensional BCS Hamiltonians with general
$l$-wave pairing interactions, classifying the models in this family that are Bethe-ansatz
solvable in the finite-size regime. We show that these solutions are characterized by
nontrivial winding numbers, associated with topological phases, in some part of the corresponding phase diagrams. By means
of a comparative study, we demonstrate benefits and limitations of
the mean-field approximation, which is the standard approach in the limit of a large number of particles. 
The mean-field analysis also allows to extend part of the results beyond integrability, clarifying 
the peculiarities associable with the integrability itself.
\end{abstract}
\maketitle

\section{Introduction}
Superconductivity, a phenomenon that is typical in condensed matter physics, but also relevant in nuclear and subnuclear physics (see, for instance, \cite{alford2008,anglani2014}), takes its origin from pairing between
fermions. It is typically described assuming an interacting (pairing) Hamiltonian
and solving it via the mean-field (MF) approximation \cite{BCS57}, 
which explicitly violates particle number conservation.
While this limitation has a small effect
on macroscopic systems, it can lead to dramatic deviations when fluctuations
are important, i.e. when dealing with a fixed small number of particles.
This justifies the interest in the study of exactly solvable models
that avoid any approximation, at the price of assuming specific forms of the interactions, 
like in the so-called Richardson model \cite{Richardson63} with $s$-wave pairing ($l=0$). 
This model is known to be integrable and its exact solution is known to be related to the Gaudin spin Hamiltonians \cite{ref1,ref2}. 
This exact-solution approach allowed various generalizations of the Richardson-Gaudin models 
\cite{germanrev,sierralungo,ortiz2014}, relevant for condensed matter and nuclear physics. 
In general, Richardson-Gaudin  particle-conserving integrable models can be classified into rational, hyperbolic, and trigonometric 
classes. Within this classification, a realization of the hyperbolic model 
is the $p_x+i p_y$ model, which has been extensively studied 
\cite{ref4,sierra09,VanRaemdonck2014,ortiz2005,ortiz2010,dukelsky2011,pan1998}, 
also in the presence of interfaces with normal conductors (see e.g. \cite{fisher2012,giuliano2013,affleck2014}).

These examples motivate the need for analyzing integrable models for superconductivity, by elucidating 
the physics of some delicate aspects of strongly correlated quantum
systems (see also \cite{recati}).  Particularly intriguing is the possibility to include pairing interactions with 
higher angular momentum (a pivotal example being the $d$-wave, i.e. $l=2$, even chiral) in two-dimensional (2D) systems, due to their direct implication for 
high-temperature superconductivity \cite{leerev}. Among the plethora of compounds and lattice schemes belonging to this family, we report  the very recent
realization of high-temperature (and likely $d$-wave) superconductivity on twisted bilayer graphene \cite{jarillo2018}.
Still on the experimental side, the $p$-wave ($l=1$) pairing is present in $^{3}$He \cite{volovik} and in strontium ruthenates \cite{mackanzie2003,maeno2012}, while
$f$-wave pairing occurs for instance in superfluid ${}^3$He \cite{gould1986,halperin2006}.
Moreover, new progress in the physics of ultracold Fermi gases opens up 
the possibility to design superconductive pairings up to the $h$-wave ($l = 5$), see e.g.  \cite{anna,mathey2007,dutta2010,sarma2010,mao2011,hao2013,wu2013,bou2017}.

Motivated by these possibilities and by the considerable theoretical interest in the high-wave superconductivity,
in the present paper we analyze a large family of 2D BCS models with arbitrary $l$-wave ($l_x + i \, l_y$) pairing interaction.  A particular attention is posed on the phase content of these models.
We first discuss (Sec. \ref{exs})  the cases that can be exactly solved via the Bethe-ansatz in a finite-size system. Later, we 
describe a standard MF analysis (Sec. \ref{MF}), and we compare the results from the two different approaches  studying
the topological properties of their solutions (Sec. \ref{winding}).
In this way, further insight is also achieved  for the cases where integrability does not hold, as well as for the role of integrability itself. 

The family of superconductive models that we are going to study is described by Hamiltonians of the form
\begin{equation}
H=\sum_{{\bf k}}\epsilon_{{\bf k}}c_{{\bf k}}^{\dagger}c_{{\bf k}}-g\sum_{{\bf kk}'}(k_{x}-ik_{y})^{l}(k'_{x}+ik'_{y})^{l}c_{{\bf k}}^{\dagger}c_{-{\bf k}}^{\dagger}c_{-{\bf k}'}c_{{\bf k}'} \, .
\label{Ham}
\end{equation}
There $c_{{\bf k}}^{\dagger}$ is the creation operators of 2D fermions
with momentum ${\bf k}=(k_x,k_y)$, and $g$ is the coupling constant, positive
for an attractive interaction.  Notice that the interaction term creates and annihilates pairs of fermions with opposite momentum.
In order to keep the widest generality, at the beginning of our analysis we do not adopt 
any particular choice for the single particle energy $\epsilon_{\b{k}}$,
only assuming it is a function of the modulus $k \equiv|\b{k}|$.

In Eq. \eqref{Ham}, we have dropped the spin index  $\{\uparrow,\downarrow\}$ in the Fermi operators, so spinless fermions are formally considered. If instead the Cooper pairs are spinful, the symmetry of their spin wavefunctions is univocally determined by the Fermi-Dirac statistics. 
In fact, when $l$ is even, the Cooper pairs form a spin singlet (antisymmetric), while when $l$ is odd they are in the triplet sector (symmetric and polarized). 
In both the cases, the structure of the Bethe-ansatz equations  and of the spatial part of the exact Cooper wavefunctions (introduced in Sect. \ref{exs}) in the presence of integrability are the same as in the spinless model described in Eq. \eqref{Ham}.  

The familiar $s$-wave case corresponds to $l=0$ and to the singlet sector of the spin wavefunction. This is the sole non symmetry-breaking case
under parity and time reversal transformation.
The breaking of these symmetries for $l\geq 1$  leads to different kinds of exact solutions, introducing nontrivial topological properties of the paired
states (according to the ten-fold way classification for the topological insulators and superconductors, see e. g.  \cite{zirnbauer1996,zirnbauer1997,ludwig2009,ludwig2010}).

\section{Exact solution in the integrable cases}
\label{exs}

\subsection{General setting}

In the present Section we address the exact solution of the Hamiltonian in Eq. (\ref{Ham}).
We find that the precise forms of $\epsilon_{\b{k}}$ and of the Cooper wavefunctions are constrained  by 
requiring the integrability.
 
The first step to proceed on is to notice that when only a single fermion
occupies the level in ${\bf k}$ or $-{\bf k}$ (i.e. without its partner), it
decouples  from the ground-state dynamics,  due to the interaction in Eq. \eqref{Ham}.
So, it is convenient to restrict ourselves to the dynamics of the Cooper pairs, having creation operators 
$b_{{\bf k}}^{\dagger}=c_{{\bf k}}^{\dagger}c_{-{\bf k}}^{\dagger}$ (see e.g. \cite{germanrev}).
Accordingly, the Hamiltonian in Eq. (\ref{Ham}) takes the form 
\beq
H=\sum_{{\bf k}}  2 \, \epsilon_{{\bf k}}\, b_{{\bf k}}^{\dagger}b_{{\bf k}}-g \, B_{0}^{\dagger}B_{0} \, .
\label{Hquad}
\eeq
Due to the particular factorized form of the interaction in Eq. (\ref{Ham}), $H$ is now quadratic
in terms of the new operator $B_{0}^{\dagger}=\sum_{{\bf k}} z_{{\bf k}} \, b_{{\bf k}}^{\dagger}$ 
where $z_{{\bf k}}=(k_{x}-ik_{y})^{l}$ are called pairing functions.
Clearly, if the $b_{{\bf k}}$ operators were truly bosonic, the Hamiltonian would
be directly diagonalizable. However, the $b_{{\bf k }}$ are instead hard-core bosons 
obeying the following commutation relations 
\beq
\left[b_{{\bf k}},b_{{\bf k}'}^{\dagger}\right]=\delta_{{\bf kk}'}(1-2 \, b_{{\bf k}}^{\dagger} b_{{\bf k}}) \, .
\label{commrel}
\eeq
 As a trial wave function for $p$ pairs, we take the following general ansatz
\begin{equation}
|\Psi_{p}\rangle=\prod_{\nu=1}^{p}B_{J_{\nu}}^{\dagger}|0\rangle \, , \qquad B_{J}^{\dagger}=\sum_{{\bf k}}w_{{\bf k}}(J) \, b_{{\bf k}}^{\dagger} \, .
\label{eq:ansatz}
\end{equation}
and we impose the eigenvalue equation 
\begin{equation}
(H-\mathcal{E}_{p})|\Psi_{p}\rangle=0 \, , 
\label{eigenv}
\end{equation}
where the total energy $\mathcal{E}_{p}$ is  given by the sum of the pair energies, 
$\mathcal{E}_{p}=\sum_{\nu=1}^{p}E_{J_{\nu}}$.

The next two sections will be devoted to the solution of Eq. (\ref{eigenv}) for one single pair and for multi-pair configurations.
Generally, these solutions are obtained using the algebra of the pseudo-bosonic commutation relations to shift $H$  
in Eq. \eqref{eigenv} through the operators $B_{J_{\nu}}^{\dagger}$ contained in $|\Psi_{p}\rangle$,
until $H$ acts on the vacuum $|0\rangle$, giving zero \cite{vondelft1999}. 
As the detailed calculation is rather cumbersome, it is presented in Appendix A.

\subsection{One pair case}
By restricting the eigenvalue equation in Eq. (\ref{eigenv}) to one pair $|\Psi_{1}\rangle$
with energy $E_{J}$, we obtain the condition:
\begin{equation}
w_{{\bf k}}(J)=g\frac{z_{{\bf k}}}{2\epsilon_{{\bf k}}-E_{J}}\sum_{{\bf k}'}z_{{\bf k}'}^{*} \, w_{{\bf k}'}(J) \, .
\label{eq:eigval cond}
\end{equation}
Multiplying both sides by $z_{{\bf k}}^{*}$ and summing in ${\bf k}$ (which is customary for
the gap equations in the BCS theory \cite{grosso,annett}), unless the "order parameter" $W(J)=\sum_{{\bf k}} \, z_{{\bf k}}^{*} w_{{\bf k}}(J)$
is zero, we obtain the Richardson equation for one pair,
\beq
1-g\sum_{{\bf k}}\frac{|z_{{\bf k}}|^{2}}{2\epsilon_{{\bf k}}-E_{J}}=0 \,  ,
\label{eq1}
\eeq
as well as the expressions for the ansatz's coefficients
\beq
w_{{\bf k}}(J)= gW(J) \, \frac{z_{{\bf k}}}{2\epsilon_{{\bf k}}-E_{J}}  \, ,
\label{exw}
\eeq
proportional to the wavefunction $\frac{z_{{\bf k}}}{2\epsilon_{{\bf k}}-E_{J}}$.
The proportionality factors $g\, W(J)$ do not depend on ${\bf k}$, thus they are irrelevant
and can be neglected, as they affect only normalizations and global phases. 
Consequently,  
without any loss of generality,  we can retain the wave function 
\beq
w_{{\bf k}}(J)=\frac{z_{{\bf k}}}{2\epsilon_{{\bf k}}-E_{J}}  \, .
\label{wavef}
\eeq

Notice that the spatial wavefunction \eqref{wavef} has the same parity of $l$ under the transformation $\bf{k}\to -\bf{k}$. 
This fact has a direct consequence on the symmetry of the spin part of the wavefunction, as discussed in the Introduction.
Moreover, if two spins $\{\uparrow,\downarrow\}$ are involved in the Cooper pair, still at fixed $l$,  the forms of the Hamiltonian in Eq. \eqref{Hquad} and of the commutators in Eq. \eqref{commrel} (as well as  of the consequent ones including the operators $B_J$, see the Appendix A) remain unchanged. Therefore,  the structure of the Bethe-ansatz equations and of the spatial part of the exact Cooper wavefunctions also do not change.

\subsection{Many pairs}

Similar to the one-pair case in the previous subsection, the ansatz in Eq. (\ref{eq:ansatz}) for the $p$ pair case reads
\begin{equation}
|\Psi_{p}\rangle=\prod_{\nu=1}^{p}B_{J_{\nu}}^{\dagger}|0\rangle \, ,\qquad B_{J}^{\dagger}=\sum_{{\bf k}}\frac{z_{{\bf k}}}{2\epsilon_{{\bf k}}-E_{J}} b_{{\bf k}}^{\dagger} \, ,
\label{eq:ansatz2}
\end{equation}
where we have assumed the expression in Eq. (\ref{wavef}) for the wavefunctions.
The solution of Eq. (\ref{eigenv}),  discussed in detail in Appendix A, yields the following final equations analogous to Eq. (\ref{eq1}). These solutions can be classified into three groups,
depending on the form of $z_{{\bf k}}$: 
\begin{enumerate}

\item The pairing function $z_{{\bf k}}$ is independent of \textbf{$\mathbf{k}$}. 
A relevant case is obtained by fixing $z_{{\bf k}}=1$; therefore, from Eq. (\ref{eq:general Richardson}), we get
the well-known Richardson equations 
\begin{equation}
1-g\sum_{{\bf k}}\frac{1}{2\epsilon_{{\bf k}}-E_{J_{\nu}}}+2g\sum_{\mu=1(\neq\nu)}^{p}\frac{1}{E_{J_{\mu}}-E_{J_{\nu}}}=0 \, ,
\label{eq:usual Richardson}
\end{equation}
whose solutions give the pair energies $E_{J_{\nu}}$ \cite{germanrev}. 
It is important to observe that here we have not imposed any restrictions on $\epsilon_{{\bf k}}$;
thus any dispersion relation (including the flat band $\epsilon_{{\bf k}}=0$) allows integrability in this case. 

\item In addition to the original $s$-wave case $z_{{\bf k}}=1$, we can also include  the choice
$z_{{\bf k}}=\exp[i\phi({\bf k})]$, where $\phi({\bf k})$ is a real function of momentum. 
Like in the previous case, the energy solutions are given by Eq. (\ref{eq:usual Richardson}),
and again there are no restrictions on $\epsilon_{{\bf k}}$. The present choice,  possibly implementable in ultracold atom set-ups by laser-assisted tunneling processes \cite{anna}, extends the previous case,
allowing for possible phases with nontrivial topology (see Appendix \ref{pure}).

\item The pairing function is $z_{{\bf k}} \propto (k_{x}-ik_{y})^{l}$. 
Since in this case $|z_{{\bf k}}|^{2}$
depends on on $\mathbf{k}$ (for $l\neq0$), we are forced to have
$|z_{{\bf k}}|^{2}\propto\epsilon_{{\bf k}}$ in order to guarantee integrability. As a consequence,
after the substitution $|z_{{\bf k}}|^{2}= \alpha \, \epsilon_{{\bf k}} = \alpha \, k^{2 l}$, Eq. (\ref{eq:general Richardson})
becomes
\begin{equation}
1-\tilde{g}\sum_{{\bf k}}\frac{\epsilon_{{\bf k}}}{2\epsilon_{{\bf k}}-E_{J_{\nu}}}
+\tilde{g} \sum_{\mu=1(\neq\nu)}^{p}\frac{E_{J_{\mu}}}{E_{J_{\mu}}-E_{J_{\nu}}}=0 \, .
\label{eq:l-wave Richardson}
\end{equation}
with $\tilde{g}=g \, \alpha$. 
For $l=1$, our result coincides with the $p$-wave solution found in \cite{sierra09}, 
with a massive-like dispersion  $\epsilon_{{\bf k}}\propto k^2$. 
Remarkably, Eq. (\ref{eq:l-wave Richardson}) also holds for the exact solution of the interesting $d$-wave case, 
where the relative angular momentum $l=2$ imposes a quartic dispersion $\epsilon_{{\bf k}}\propto  k^4$. 
\end{enumerate}

In \cite{sierra09, sierralungo} a detailed analysis was performed for the case (3),  with  $\epsilon_{\b{k}}  = k^{2n}$ and $n  = l = 1$,  both by a MF approach in the thermodynamic limit and by comparing its results  with the properties of the exact wavefunction  from the solution of the Bethe-ansatz equations. The topological aspects of the obtained phases were also discussed. 

In the following, we generalize the latter analysis to the wider situation where $n,l \geq 1$,  $l$ ($n$) is assumed to be an integer (half-integer), and $n , l$ are allowed to be different. If $l \neq n$,  integrability is broken, so that only a MF approach can be used. If, instead, $n = l$, a deeper knowledge is achieved by studying again  the topological properties of the exact wavefunctions. 

We mention finally that integrability is not spoiled if an additional constant is added to the quasiparticle dispersion $\epsilon_{\b{k}}$, as done in \cite{links2012}. There Eqs. \eqref{eq:usual Richardson} and \eqref{eq:l-wave Richardson} were written in a implicit manner. Moreover, if $n \neq l$, integrability can sometimes be preserved if additional Hamiltonian terms are added; an explicit example is given \cite{marquette2013}.

\section{mean-field analysis}
\label{MF}

\subsection{General formalism}
In this section we analyze the MF properties of the Hamiltonian in Eq. (\ref{Ham}).
Following the standard approach to MF superconductivity \cite{grosso,annett}, 
we find that the MF quadratic Hamiltonian, in the thermodynamic limit and  in the grand-canonical ensemble, 
derived from the one in Eq. (\ref{Ham}), is
\beq
H =  E_c + \sum_{\b{k} }  \Big( \xi_{k}  \,  \cop{c}_{\b{k}}  \, \aop{c}_{\b{k}} +  
\Delta (k_x + i k_y)^l  c_{\b{k}} \aop{c}_{\b{- k}} + \mathrm{H.c.} \Big), 
\label{H}
\eeq
 where $E_c$ is the condensation energy, defined below, and $\xi_{k}   = (\epsilon_{k} - \mu) = (k^{2n} - \mu)$ is the rescaled dispersion. 
In the chemical potential $\mu$,  the Hartree terms are also included, coming from the Wick contractions of 
the interaction term in the Hamiltonian of Eq. (\ref{Ham}). According to the analysis performed in Sect. \ref{exs}, the 
integrable cases correspond to $n=l$; however, for the sake of completeness, here we do {\emph not} fix $n$  and $l$ to be equal in this MF treatment.  

The Hamiltonian in Eq. \eqref{H} describes  potentially realistic cases if $n = 1$ and $l = 2$ (when two spins are considered) \cite{annett}, and if $n = l = 1 $ \cite{volovik,mackanzie2003,maeno2012}.

In  Eq. (\ref{H}) we set  $ \Delta = \sum_{\b{k}^{\prime}} \, g \, (k'_{x}+ik'_{y})^{l} \, \langle c_{-\b{k}'}c_{\b{k}'} \rangle$, with $\langle c_{-\b{k}'}c_{\b{k}'} \rangle$ being the vacuum expectation value
of the superconductive ground-state. Therefore, the gap function can be written as  $\Delta_{\mathbf{k}}  = \Delta \, (k_x + i k_y)^l$; 
the quantity $(k_x + i k_y)^l$ coincides, up to a constant, with the spherical harmonic $Y^l_l(\hat{k})$ projected in the 2D plane (expected to be the  more stable one in the absence of external strains or pressures, see e.g. \cite{annett}). 

The condensation energy $E_{\mathrm{C}}$ is given by
\beq
E_{\mathrm{C}} = - 4  \sum_{\b{k} , \b{k^{\prime}} >0}    \, 
\frac{\Delta_{\b{k}} \Delta^{*}_{\b{k}^{\prime}}}{g_{\b{k} \b{k^{\prime}}}} = A \,  \frac{M \Delta^2}{g} \, ,
\eeq
where the integer $M$ denotes the number of states in the  
 region of phase-space considered and $g_{\b{k} \b{k^{\prime}}}$ is the two-body potential 
appearing in the full Hamiltonian expressed in momentum space. 
In a general case, the quantity $A$ explicitly depends on the assumed form of $g_{\b{k} \b{k^{\prime}}}$. 
For the Hamiltonian  in Eq. (\ref{Ham}), this potential reads
\beq
g_{\b{k} \b{k^{\prime}}} = - g (k_{x}-ik_{y})^{l}(k'_{x}+ik'_{y})^{l} \, ,
\eeq
so that $A = 1$. As we will check in the following, an important feature of the ground-state free energy $F_{\mathrm{GS}}$ is that, when expressed as a sum on the momenta via the gap equation, it does not depend on $A$. 

The Bogoliubov spectrum corresponding to the Hamiltonian in Eq. (\ref{H}) is
\beq
\lambda_k = \sqrt{\xi_{k}^2 + \Delta^2 k^{2l}} 
\label{spectrum}
\eeq
(with $k$ denoting again the modulus of $k_x-i \, k_y$). This spectrum is gapless at $\mu = 0$ and $k = 0$.

The ground-state free energy $F_{\mathrm{GS}}  = E_{\mathrm{GS}}  + \mu \, N $, $N = 2 p$, corresponding to the spectrum in Eq. \eqref{spectrum}, is
\beq
F_{\mathrm{GS}} = \sum_{\b{k} >0} \, \big( \xi_{k}  - \lambda_k \big) +  \frac{M \Delta^2}{g} + \mu \, N \, ,
\eeq
independent of $A$, as anticipated.
The Bogoliubov coefficients are
\beq
|u_{k}|^2 = \frac{1}{2} \left( 1 +  \frac{\xi_{k} }{\sqrt{\xi_{k}^2 + \Delta^2 k^{2l}}} \right), \quad |v_{k}|^2 = 1 - |u_{k}|^2 \, ,\\
\eeq 
so that the MF wave function results:
\beq
w_{\b{k}}^{(\mathrm{MF})} = \frac{v_{k}}{u_{k}} = \frac{\lambda_k - \xi_{k} }{\Delta \, (k_x + i \, k_y)^l} \, .
\label{MFW}
\eeq
The equations for $\Delta$ and  $\mu$ are as follows: 
\beq
\frac{\partial F_{\mathrm{GS}}}{ \partial \Delta} = 0    \, \rightarrow \,  \frac{M}{g} = \frac{1}{2} \, \sum_{\b{k} >0}  \, \, \frac{k^{2l}}{\lambda_k}  \, ,
\label{eqDelta}
\eeq
\beq
\frac{\partial F_{\mathrm{GS}}}{ \partial \mu} = 0    \, \rightarrow \, N = \sum_{\b{k} >0}  \, \left(1- \frac{\xi_{k} }{\lambda_k} \right) \, .
\label{eqmu} 
\eeq
The last equation can also be written as
\beq
\mu \,  \sum_{\b{k} >0} \, \frac{1}{\lambda_k}  = N  + \sum_{\b{k} >0}  \, \frac{k^{2n}}{\lambda_k}  - \frac{M}{2} 
\label{eqmu2}
\eeq
which, in the case of $n = l$, becomes, from Eq. \eqref{eqDelta},
\beq
\mu \,  \sum_{\b{k} >0}  \, \frac{1}{\lambda_k}  = N +  2 \, \frac{M}{g}  -  \frac{M}{2}  \, .
\label{eqmu3}
\eeq
Using Eq. \eqref{eqDelta}, the ground-state  free energy is  written as:
\beq
F_{\mathrm{GS}} =  \sum_{\b{k} >0} \, \Bigg( \xi_{k}  - \lambda_k  +  \frac{\Delta^2}{2} \, \frac{k^{2l}}{\lambda_k}  \Bigg) + \mu \, N  \, .
\eeq
and, exploiting Eq. \eqref{eqmu2}, also as:
\beq
F_{\mathrm{GS}} = \sum_{\b{k} >0} \, k^{2 n} \, \Bigg( 1  -  \frac{2 k^{2n} -2 \mu + \Delta^2\, k^{2(l-n)}}{2 \, \lambda_k}  \Bigg)   \, .
\label{ffin}
\eeq
If $n = l$, the latter expression shows a duality between different MF solutions,  in that
two solutions (labeled 1 and 2) are related by the equations $\mu_1 = -\mu_2$ and
$\Delta^2_1 - 2 \, \mu_1 = \Delta^2_2 - 2 \, \mu_2$, such that the corresponding free energies coincide:
$F_{\mathrm{GS}}^{(1)} = F_{\mathrm{GS}}^{(2)}$. If $n = l = 1$, this duality is justified by the exact solution of the Richardson equations \eqref{eq:usual Richardson}.

Once one considers working in a lattice, as opposed to the continuum, the above analysis can be extended straightforwardly. 
Some spin models are, indeed, quadratic in Fermi operators in momentum space with pair creation \cite{Campos2010}.
For sufficiently small interaction strength $\propto g$, we expect that superconductivity involves only quasiparticles with momenta within a small range $\delta k \approx g^{\frac{1}{n}}$ around the Fermi momentum $k_F$. Here the lattice dispersion, with discretized momenta, can be expanded in powers of $k$, such that it ends up in a power-law dispersion. At that point, the MF analysis proceeds as described before.

\subsection{Mean-field phase diagram}
Using the derived expressions for the ground-state free energy, for the wave functions of the Bogoliubov excitations, and for the self consistency equations, 
it is interesting to characterize the phase diagram of the Hamiltonian in Eq. (\ref{H}), as a function of $g$ and of the (average) filling $N/M \equiv x$.

Various transition lines, between different quantum phases, can be identified. 
A notable transition occurs  at $\mu = 0$, where
the spectrum in Eq. (\ref{spectrum}) is gapless at $k  = 0 $. There the MF wavefunction behaves as:
\beq
w_{\b{k}}^{(\mathrm{MF})} \approx \left\{
    \begin{array}{rl}
    (k_x-ik_y)^{l} \, k^{2(n-l)} & \text{if } \, \mu < 0 \, \, \, \text{and} \, \, \, n \geq l , \\
      (k_x-ik_y)^{l} & \text{if } \, \mu < 0 \, \, \, \text{and} \, \, \, n<l , \\
          \frac{1}{(k_x+ik_y)^{l}} & \text{if } \, \mu > 0 \, .
    \end{array} \right.
    \label{WFlim}
\eeq
This transition has a  nature similar to the Read-Green one described in the case $n=l=1$ \cite{RG,sierra09, sierralungo,ortiz2010} (and found to be a third-order transition in \cite{ortiz2010}); for this reason in the following the same name will be adopted for it. The condition $\mu = 0$ translates, from Eq. (\ref{eqmu3}), into the relation
\beq
x = \frac{1}{2} \, \left(1- \frac{4}{g} \right) \, .
\label{MRline2}
\eeq
The line identified by this equation does not depend on the distribution of the momenta, thus is topologically protected against every perturbation changing it, and possibly breaking the integrability of the  Hamiltonian in Eq. (\ref{Ham}).  

Another  notable line, denoted as the (generalized) Moore-Read line \cite{sierra09,links2015}, is found for every $n = l$, parametrized by the relation  $\mu =  \frac{\Delta^2}{4}$; along this line the condition $F_{\mathrm{GS}} = 0$ holds: the same free energy of the vacuum, intended as the absence of fermions ($x = 0$), is obtained for the superconductive ground-state. Notice that, in order to obtain this result,  the positiveness of $\mu$ is crucial.
The condition $\mu =  \frac{\Delta^2}{4}$ is fulfilled on the line:
\beq
x=  \Big(1 - \frac{4}{g} \Big) \, ,
\eeq
a result found by exploiting Eq. \eqref{eqmu}.
There the mass gap does not vanish but the ground-state free energy is discontinuous in the thermodynamic limit.

As for the case $n = l = 1$ \cite{sierra09,links2015}, the duality mentioned in the previous section holds, at least at the MF level, between a point $(g, x_w)$ in the weak pairing regime ($\mu >0$) and a point $(g, x_s)$ in the strong pairing regime ($\mu <0$); these points are related to each other by the relation 
\beq
x_w +x_s = \Big(1-\frac{4}{g}  \Big) 
\eeq
which is still obtained directly from Eqs. \eqref{eqmu} and \eqref{eqmu3}.
Therefore, the Read-Green line is self-dual, 
while the MR state is dual to the vacuum, where $x =0$. 

The Read-Green and Moore-Read lines meet at the point $g = 4$, where the limit $x = 0$ is achieved.

By a direct numerical analysis of the MF free energy in Eq. \eqref{ffin},  performed on various cases with $n \neq l$, we have found strong indications that the Moore-Read line does not persist  out of the integrability \cite{notevan}, as $F_{\mathrm{GS}} \neq 0$.

If $n =l$, the minimum  $E_{\rm{GAP}}$ of $\lambda_k$, Eq. (\ref{spectrum}), is 
\beq
E_{\rm{GAP}} =   \left\{
    \begin{array}{rl}
   |\mu|   & \text{if } \, \mu < \frac{\Delta^2}{2} \, ,\\
      \Delta \sqrt{\mu - \frac{\Delta^2}{4}}  & \text{if } \, \mu > \frac{\Delta^2}{2} \, .
    \end{array} \right.    
    \label{voline}
\eeq
 The condition $\mu = \frac{\Delta^2}{2}$ defines a third notable transition line, the so-called Volovik line \cite{sierra09,sierralungo}. Along it a first-order quantum phase transition,
 reminiscent of the Higgs transition, occurs \cite{volovik}.
 The same line depends on the distribution of the momenta, thus it is \emph{not} topologically protected (and its presence must be verified beyond the MF approach, adopted in the following). Setting $\mu = \frac{\Delta^2}{2}$ and exploiting Eqs. \eqref{eqDelta} and \eqref{eqmu3}, we find that, if $n = l$, the Volovik line reads explicitly as:
\beq
x = \frac{1}{2} \Bigg(1 - \frac{1}{M}  \,  \sum_{\b{k} >0}  \, \,  \frac{2 k^{2l}- \Delta ^2}{\lambda_k} \Bigg) \, .
\label{vol}
\eeq
 From a numerical study of $\lambda_k$ in Eq. (\ref{spectrum}), we conclude that the Volovik line does not survive  if $n<l$, since $E_{\rm{GAP}}$ always arises at $k \neq 0$. On the contrary, if $n>l$, $E_{\rm{GAP}}$ is located at $k = 0$ for some values of $\Delta$ and $\mu$, so that a Volovik line can still be identified  (the defining equation, similar to (\ref{vol}), is not easily writable as a closed formula).

\section{Topological properties}
\label{winding}
In this section we give a deeper characterization of the MF phase diagram, sketched in the previous section, studying the  topology of the various identified phases. Focusing first on the case $n = l$, we start by taking the MF Cooper wavefunction $w_{\b{k}}^{(\mathrm{MF})}$ in Eq. \eqref{MFW} to calculate the topological invariant \cite{sierralungo}:
\beq
I_{\mathrm{MF}}  = \frac{1}{4 \pi} \, \int_{S^2}   \mathrm{d} \b{k} \, \, w_{{\bf k}} \, ,
\label{defI}
\eeq
where $S^2$ is the sphere of radius $|\b{k}| = 1$ obtained from the plane $R^2$
 by the inverse of the stereographic projection \cite{bernevig,noteint}.
We obtain  $I_{\mathrm{MF}} = l$ if $\mu >0$, and $I_{\mathrm{MF}} = 0$ if $\mu <0$. 
This result matches the previously found values $I_{\mathrm{MF}} = 1$ for the $p$-wave case \cite{RG,sierra09,sierralungo} and  $I_{\mathrm{MF}} = 2$ for the $d$-wave case \cite{RG}.
As generally expected (see e. g. \cite{ludwig2009}), $I_{\mathrm{MF}}$  is sensitive to the vanishing of the energy for the Bogoliubov quasiparticles, occurring at $\mu = 0$. Finally, it is worth noticing that, although the location of the Read-Green line is independent of the momentum distribution and of the Bogoliubov dispersion law  $\lambda_{k}$, the (topological) phases bounded by it depend on $l$. This index can affect the topology since it induces global (on the entire set of the allowed momenta) and not smooth ($l$ is discrete) modifications on $\lambda_{k}$.

The topological content of the phase diagram can be inferred, not only from the MF wavefunction of a single Cooper pair, Eq. \eqref{MFW}, but also from the MF ground-state wavefunction,  
following a procedure common in the study of topological insulators and superconductors \cite{bernevig}.  In particular, denoting by $|u_{\bf{k}} \rangle$ the positive-energy eigenvector of the quadratic Hamiltonian in Eq. \eqref{H}, $I_{\mathrm{MF}}$ is expressed as the integral on the momentum space of the Berry curvature:
\beq
I_{\mathrm{MF}} = \frac{1}{4 \, \pi}  \, \int_{S^2}  \mathrm{d} \b{k} \, \,  \b{\nabla}_{\b{k}} \times  \langle u_{\b{k}} | \b{\nabla}_{\b{k}} | u_{\b{k}} \rangle \, .
\label{MWinv}
\eeq 
The equivalence between  the two MF calculations for $I_{\mathrm{MF}}$
stems directly from the fact that  $|u_{\bf{k}} \rangle$ is an excited state obtained by breaking a Cooper pair. In turn, the expression \eqref{MWinv} is also equivalent to the spin-texture one
\cite{bernevig,RG,anderson1958}
 \beq
I_{\mathrm{MF}} = \frac{1}{8 \, \pi}   \, \int_{S^2}  \mathrm{d} \b{k} \, \, \epsilon_{abc} \epsilon_{ij} \,  \hat{d}_a (\b{k}) \, \partial_{k_i}  
\hat{d}_b (\b{k}) \, \partial_{k_j} \hat{d}_c (\b{k})  
\label{MWinv2}
\eeq
\big($(i,j) = {\{x,y\}}$, and $(a,b,c) = \{1, 2, 3\}$\big), obtained expressing the
 Hamiltonian \eqref{H} in terms of the Pauli matrices, in the basis $\big(c_{\b{k}}, c_{-\b{k}} \big)^T$:
$H = \sum_{\b{k}} \hat{d}(\b{k}) \cdot \b{\sigma}$.
Direct numerical calculation of both the expressions \eqref{MWinv} and \eqref{MWinv2} confirmed the result $I_{\mathrm{MF}} = l$
if $\mu >0$.
\begin{figure}[h!]
\includegraphics[scale=0.20]{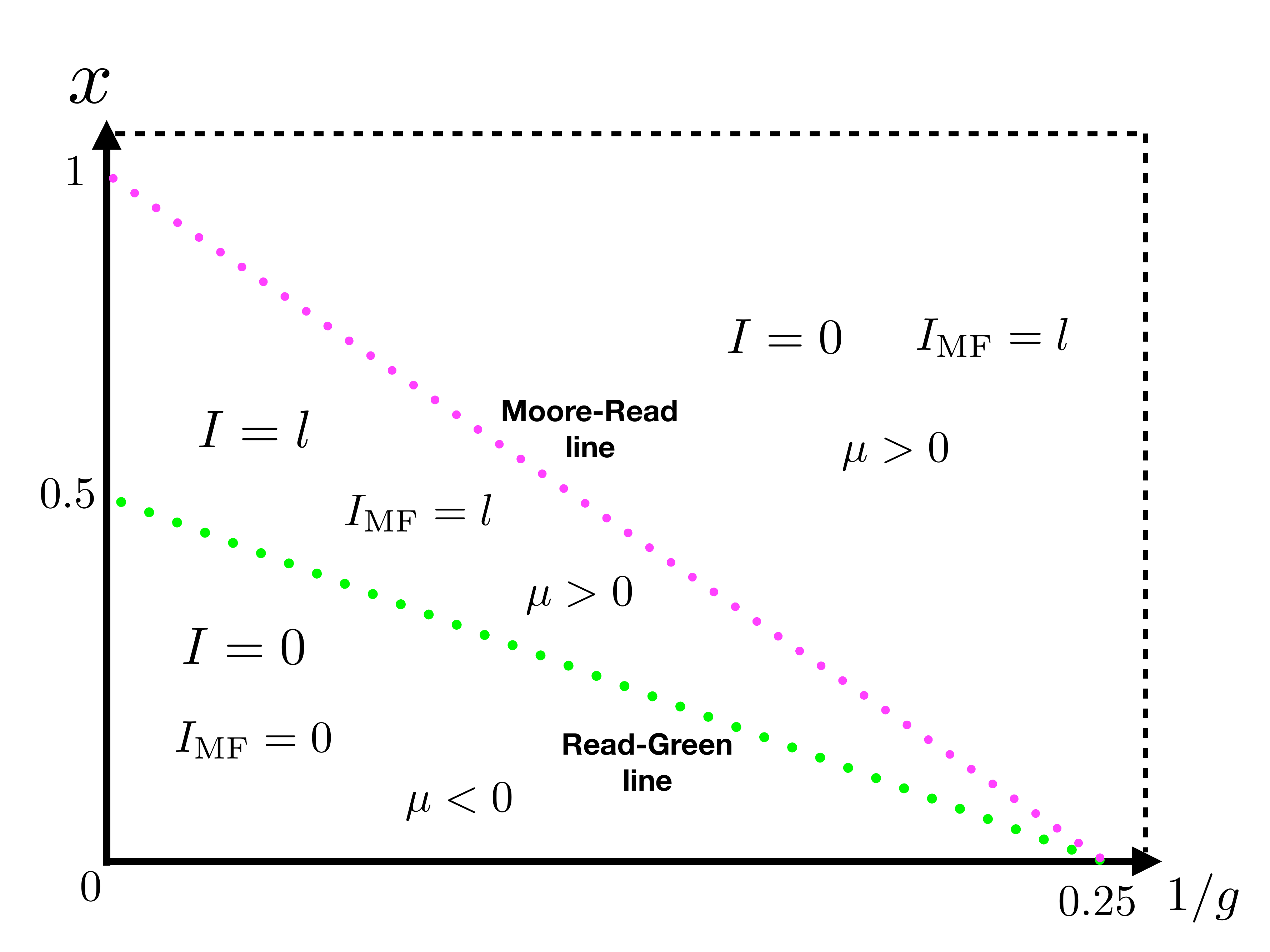}
\caption{Mean-field phase diagram for $n =l$, as a function of $x$ and $g$. The topological invariant  $I_{\mathrm{MF}}$, relative to a single Cooper pair,  is reported, as well as the invariant $I$ from the exact wavefunction in Eq. \eqref{psi}. Notice the difference between MF and exact invariants in the phases above the Moore-Read line. The different length scale for the axes is chosen for sake of clarity of the picture. 
The Moore-Read line disappears in general out of integrability, if $n \neq l$.}
\label{grafico}
\end{figure}

The content in topology obtained using the MF wavefunctions can be probed  
also calculating the same quantity as in Eq. \eqref{defI} in terms of the exact wavefunction $w_{\b{k}}$ of a single Cooper pair, then considering again the limit $x = 0$. We implicitly assume that fluctuations beyond MF do not change the MF phase diagram significantly; thus the solution of the Bethe-ansatz equations essentially leads to  the same phase diagram. This hypothesis will not be contradicted in the following. The exact wavefunction, derived in Section \ref{exs}, reads, up to an unimportant multiplicative constant:
\beq
w_{{\bf k}}=\frac{(k_x - i k_y)^l}{2\epsilon_{{\bf k}}-E}   \, ,
\label{psi}
\eeq
where $E$ is the pair energy (complex in general \cite{germanrev}), derived from the solution of the Richardson equations.  
The integral as in Eq. \eqref{defI} can be recast as follows:
\beq
I =  l^2 \, \int_0^{\infty} \mathrm{d} u \, \frac{ u^{(3l-1)} - E \bar{E} \, u^{(l-1) }}{(u^l+ (u^l -E)(u^l -\bar{E}))^2} \, ,
\label{wind}
\eeq
with $u  = k^2$. The result of Eq. (\ref{wind}) is 
\beq
\left\{
\begin{array}{c}
I= l \, \, \, \mathrm{if} \, \, \, E= 0\\ 
{}\\
I = 0  \, \, \, \mathrm{if} \, \, \, E\neq 0 \, .
\end{array} \, \right.
\label{Iex}
\eeq 
An alternative derivation of the winding number $I$ is discussed in the Appendix \ref{alternativeI}; this turns out to be useful also for the pure phase case in the Appendix \ref{pure}.  Moreover, it would also be interesting  to extend the calculation of $I$ to multi-pairs states, e. g., following the approaches in \cite{marquette2013,foster2013}.

Referring to the MF diagram in Fig. \ref{grafico}, the condition $E = 0$ in \eqref{Iex} is fulfilled if $x = 0$, at the intersection with the Moore-Read line, where $g = 4$.  This fact indicates that $I = l$ in the region between the Read-Green and the Moore-Read lines, while $I = 0$ in the other phases.
Therefore, $I$  matches the MF phase diagram, oppositely to $I_{\mathrm{MF}}$: indeed  $I_{\mathrm{MF}}$ is  nonvanishing also in the region to the right of the Moore-Read line, thus  $I_{\mathrm{MF}}$ it does not detect this line. The described mismatch is indeed interesting, since it can indicate a general inability of the topological invariants from the MF wavefunctions to correctly detect some phases of (topological) insulators or superconductors. In our case, the mismatch occurs since the mass gap does not vanish on the MR line. It remains an open question whether the origin of the puzzle is due to integrability of the full model in Eq. \eqref{Ham}. However, such interpretation is suggested by the fact that from the MF analysis the MR line seems generally absent for $n \neq l$, where integrability is broken (and no divergencies occur in the spectrum, a situation found instead in the presence of long-range Hamiltonian couplings, see \cite{lepori2016} and references therein \cite{lepori2017}).

We note finally that in \cite{ortiz2010} it has been suggested, for the case $n = l =1$, that the Moore-Read line does not identify a genuine quantum
phase transition, a possibility partly solving the mismatch mentioned above. However, the same result for $I$ (different from zero only at $E = 0$) from the exact pair wavefunction, in \cite{sierralungo} and in the  present paper, seems to rule out this scenario.

\section{Discussion and conclusions}
\label{concl}
In this paper we have analyzed the physical features of a large set of superconductive models for which an exact solution is available,
composed of two-dimensional systems with a factorized form for the momentum dependent interaction. Besides the 
known cases of the $s$-wave pairing, solved by Richardson \cite{Richardson63}, and $p$-wave pairing, discussed for the first time by 
Ibañez \textit{et al.} in \cite{sierra09}, we have found that, in general, $l$-wave pairing is exactly solvable on a finite-size system, provided that the single particle dispersion is proportional to $k^{2n}$, with $n = l$.

Analyzing the integrable cases, we also found that  the topological invariants calculated in the framework of the mean-field approach can not  reproduce correctly the phase diagrams of the considered integrable models, in contrast to the corresponding invariants obtained from the exact (Bethe-ansatz) solutions. This discussion has shown the potential inadequacy of the mean-field topological invariants to predict the correct phase diagram of (topological) insulators and superconductors, at least in peculiar situations. In our case, the origin of this problem seems to be the (possible) presence of quantum phase transitions without vanishing of the mass gap, a feature possibly related to integrability. We notice that quite recently a change in topology without mass gap closing, in the presence of large interaction, was found numerically in \cite{capone2015}.

In the non-integrable cases $n \neq l$ (as well as for other perturbed models where interactions do not assume the special form of Eq. \eqref{Ham}), the exact wavefunctions analogous to Eq. \eqref{wavef} cannot be derived, because the Bethe-ansatz is not applicable, therefore only the the mean-field approach can be exploited.  The reliability of this approach out of the integrable regime is suggested also by its prediction about the general absence of quantum phase transitions with nonvanishing mass gap.\\

{\bf Acknowledgements --}
The authors are pleased to thank Miguel Ibanez Berganza, Simone Paganelli, and German Sierra for useful discussions.

\vspace{2cm}

\appendix

\section{Bethe-ansatz solution of equation \eqref{eigenv}}

The eigenvalue equation in (\ref{eigenv}) can be written as
\begin{equation}
\left(\left[H,\prod_{\nu=1}^{p}B_{J_{\nu}}^{\dagger}\right]-\mathcal{E}_{p}\prod_{\nu=1}^{p}B_{J_{\nu}}^{\dagger}\right)|0\rangle=0,\label{eq:eigenv-bis}
\end{equation}
where the commutator on the left side expands as
\begin{equation}
\sum_{\nu=1}^ {p}\left\{ \left(\prod_{\eta=1}^{\nu-1}B_{J_{\eta}}^{\dagger}\right)\left[H,B_{J_{\nu}}^{\dagger}\right]\left(\prod_{\mu=\nu+1}^ {p}B_{J_{\mu}}^{\dagger}\right)\right\} .\label{eq:comm_H_PB}
\end{equation}
Using the relations
\begin{align}
\left[b_{\mathbf{k}}^{\dagger}b_{\mathbf{k}},B_{J}^{\dagger}\right] & =w_{\mathbf{k}}(J)b_{\mathbf{k}}^{\dagger},\\ \nonumber
\left[B_{0},B_{J}^{\dagger}\right] & =\sum_{\mathbf{k}}z_{\mathbf{k}}^{*}w_{\mathbf{k}}(J)(1-2b_{\mathbf{k}}^{\dagger}b_{\mathbf{k}})
\label{comm}
\end{align}
we find the expression for every single commutator appearing in Eq. (\ref{eq:comm_H_PB}):
\begin{eqnarray}
\left[H,B_{J}^{\dagger}\right] & = & E_{J}B_{J}^{\dagger}+\sum_{\mathbf{k}}(2\epsilon_{\mathbf{k}}-E_{J})w_{\mathbf{k}}(J)b_{\mathbf{k}}^{\dagger}\nonumber \\
 &  & -gB_{0}^{\dagger}\sum_{\mathbf{k}}z_{\mathbf{k}}^{*}w_{\mathbf{k}}(J)(1-2b_{\mathbf{k}}^{\dagger}b_{\mathbf{k}}) \, .
 \label{eq:comm_H_B}
\end{eqnarray}
Putting Eq. (\ref{eq:comm_H_B}) in Eq. (\ref{eq:comm_H_PB}) and using the basic relation
$H|0\rangle=0$, we find:
\begin{widetext}
\begin{eqnarray}
H|\Psi_ {p}\rangle & = & \mathcal{E}_ {p}|\Psi_ {p}\rangle+\sum_{\nu=1}^ {p}\left[\left(\sum_{\mathbf{k}}(2\epsilon_{\mathbf{k}}-E_{J_{\nu}})w_{\mathbf{k}}(J_{\nu})b_{\mathbf{k}}^{\dagger}-gB_{0}^{\dagger}\sum_{\mathbf{k}}z_{\mathbf{k}}^{*}w_{\mathbf{k}}(J_{\nu})\right)\left(\prod_{{\eta=1\atop \eta \neq\nu}}^ {p}B_{J_{\eta}}^{\dagger}\right)\right]|0\rangle\nonumber \\
 &  & +\sum_{\nu=1}^ {p}\left\{ \left(\prod_{\eta=1}^{\nu-1}B_{J_{\eta}}^{\dagger}\right)2gB_{0}^{\dagger}\left(\sum_{\mathbf{k}}z_{\mathbf{k}}^{*}w_{\mathbf{k}}(J_{\nu})b_{\mathbf{k}}^{\dagger}b_{\mathbf{k}}\right)\left(\prod_{\mu=\nu+1}^ {p}B_{J_{\mu}}^{\dagger}\right)\right\} |0 \rangle \, .
\label{eq:H_x_psi}
\end{eqnarray}
In the last term of Eq. (\ref{eq:H_x_psi}), we want to commute the operator
$b_{\mathbf{k}}^{\dagger}b_{\mathbf{k}}$ to the extreme right, where it annihilates
the vacuum $|0 \rangle$. To this aim, we write this term as
\begin{equation}
\sum_{\nu=1}^{p}\left\{ 2gB_{0}^{\dagger}\left(\prod_{\eta =1}^{\nu-1}B_{J_{\eta}}^{\dagger}\right)\sum_{\mu=\nu+1}^{p}\left\{ \left(\prod_{\eta'=\nu+1}^{\mu-1}B_{J_{\eta'}}^{\dagger}\right)\left[\sum_{\mathbf{k}}z_{\mathbf{k}}^{*}w_{\mathbf{k}}(J_{\nu})b_{\mathbf{k}}^{\dagger}b_{\mathbf{k}},B_{J_{\mu}}^{\dagger}\right]\left(\prod_{\mu'=\mu+1}^{p}B_{J_{\mu'}}^{\dagger}\right)\right\} \right\} |0\rangle.
\label{eq:last_term}
\end{equation}
\end{widetext}
At this point, it is crucial to use the following manageable form
for the commutator in Eq. (\ref{eq:last_term}):
\begin{equation}
\left[\sum_{\mathbf{k}}z_{\mathbf{k}}^{*}\, w_{\mathbf{k}}(J_{\nu}) \, b_{\mathbf{k}}^{\dagger}b_{\mathbf{k}},B_{J_{\mu}}^{\dagger}\right]=\sum_{\mathbf{k}} z_{\mathbf{k}}^{*} \, w_{\mathbf{k}}(J_{\nu})w_{\mathbf{k}}(J_{\mu}) \, b_{\mathbf{k}}^{\dagger} \, .
\label{eq:comm_critical}
\end{equation}
In general, for every $\mu$ and $\nu$, we want to express Eq. (\ref{eq:comm_critical})
in the form $C_{\mu,\nu}B_{J_{\nu}}^{\dagger}+D_{\mu,\nu}B_{J_{\mu}}^{\dagger}$,
where $C_{\mu,\nu}$ and $D_{\mu,\nu}$ are some coefficients.  For this reason, we impose the condition
\begin{equation}
\sum_{\mathbf{k}}z_{\mathbf{k}}^{*}w_{\mathbf{k}}(J_{\nu})w_{\mathbf{k}}(J_{\mu})b_{\mathbf{k}}^{\dagger}=C_{\mu,\nu}B_{J_{\nu}}^{\dagger}+C_{\nu,\mu}B_{J_{\mu}}^{\dagger} \, ,
\label{eq:condition}
\end{equation}
where we have used the symmetry under the exchange $\nu\leftrightarrow\mu$.
Assuming that Eq. (\ref{eq:condition}) is correct, then we find that
the eigenvalue equation (\ref{eq:H_x_psi}) holds, provided that
\begin{equation}
1-g\sum_{\mathbf{k}}\frac{|z_{\mathbf{k}}|^{2}}{2\epsilon_{\mathbf{k}}-E_{J_{\nu}}}+2g\sum_{\mu=1(\neq\nu)}^ {p}C_{\nu,\mu}=0 \, ,
\label{eq:general Richardson}
\end{equation}
where we have used the expression for the wave function $w_{\mathbf{k}}(J)=\frac{z_{\mathbf{k}}}{2\epsilon_{\mathbf{k}}-E_{J}}$.
Equation (\ref{eq:condition}) gives 
\[
(2\epsilon_{\mathbf{k}}-E_{J_{\mu}})C_{\mu,\nu}+(2\epsilon_{\mathbf{k}}-E_{J_{\nu}})C_{\nu,\mu}=|z_{\mathbf{k}}|^{2}
\]
with two different kind of solutions: 
\begin{enumerate}
\item \textbf{\emph{s}}\textbf{-wave}. In this case $|z_{\mathbf{k}}|^{2}=1$ and
$C_{\mu,\nu}=-C_{\nu,\mu}=(E_{J_{\nu}}-E_{J_{\mu}})^{-1}$. Thus,
from (\ref{eq:general Richardson}) we get the well-known Richardson
equation Eq. (\ref{eq:usual Richardson}), with no restrictions on
$\epsilon_{\mathbf{k}}$. Notice that the condition $|z_{\mathbf{k}}|^{2}=1$ is more
general than the $s$-wave case $z_{\mathbf{k}}=1$. 
\item \textbf{\emph{l}}\textbf{-wave}. In this case, $z_{\mathbf{k}}=(k_{x}-ik_{y})^{l}$
depends on $\mathbf{k}$ (for $l\neq0$) and the coefficients are
given by 
\begin{equation}
C_{\mu,\nu}=\frac{|z_{\mathbf{k}}|^{2}}{2\epsilon_{\mathbf{k}}}\frac{E_{J_{\nu}}}{E_{J_{\nu}}-E_{J_{\mu}}} \, ,
\label{eq:coeff solution}
\end{equation}
but we must have $|z_{\mathbf{k}}|^{2}\propto\epsilon_{\mathbf{k}}$ to have an $C_{\mu,\nu}$
independent of $\mathbf{k}$. As a consequence, after the substitution $|z_{\mathbf{k}}|^{2}=\alpha \, \epsilon_{\mathbf{k}}$,
Eq.(\ref{eq:general Richardson}) becomes Eq. (\ref{eq:l-wave Richardson}).
\end{enumerate}

\section{Alternative calculation of $I$}
\label{alternativeI}

In this appendix we discus an alternative derivation of the winding number $I$, which is also useful for the pure phase case in the Appendix \ref{pure},
that can be performed by analyzing directly the map $\omega_{\mathbf{k}}$ in the case of real $E$. In order to do that, we  first separate Eq. (\ref{psi}) as
\beq
\omega_{\mathbf{k}} = \big(f_- (k) + f_+ (k) \big) \, e^{i \phi_k l} 
\eeq
with $ f_- (k) =  \frac{k^l}{k^{2l}- E }$, $k < E^{1/2l}$, and $ f_+ (k) =  \frac{k^l}{k^{2l}- E}  $, $k > E^{1/2l}$.\\
The part  $ f_+ (k)  \, e^{i \phi_k l}$ gives a contribution $I_+ = l$ to $I$, since $ f_+ (k)$ is monotonic in $k$ and  assumes values $[0, \infty)$, so that $f_+ (k) \,  e^{i \phi_k l}$
covers $l$ times (because of the phase $l \, \phi_k$)  the entire plane $R^2 \sim S^2$ (the identification relying again on the stereographic projection).\\
Assuming  now that $E \neq 0$, we put $k = 1/p$ in  $ f_- (k)$ , obtaining
$ f_- (p) =- \frac{1}{E} \, \frac{p^l}{p^{2l} - E} = - f_+(p)$, with $p > E^{1/2l}$.
Apart from the unimportant multiplicative factor $E^{-1}$, we can write (renaming $p \equiv k$)
\beq
\omega_{\mathbf{k}} = \big(f_- (k) - f_- (k) \big) \, e^{i \phi_k l} \, = 0 \,
\eeq
showing that $I = 0$ if $E \neq 0$. The minus sign in $ f_- (p)$, responsible for the vanishing result for $I$, is related to the fact that, for $k$ varying,   $ f_+ (k)$
and  $ f_- (k)$ span the space $R^2 \sim S^2$ in the opposite sense.\\
The situation is different if $E = 0$: in this case we get only
\beq
\omega_{\mathbf{k}} =  f_+ (k)  \, e^{i \phi_k l} 
\eeq
and $I = I_+ = l$.

\section{Pure phase gap}
\label{pure}

We can also calculate the topological index $I$ in the case when  $\Delta(k) = e^{i \phi_{\hat{k}} l}$. In this case, we have shown in Sec. \ref{exs} that we have integrability, no matter what  the particular single particle dispersion  $\epsilon_{\mathbf{k}}$ is; therefore, we assume again $\xi_{(l)} (k) = k^{2l}$.
 The exact wave function reads, in momentum space and up to an unimportant multiplicative constant,
\beq
\omega_{\mathbf{k}} = \frac{(k_x - i k_y)^l}{k^l \, \big( 2\epsilon_{\mathbf{k}}- E \big)}   \, ,
\label{psimod}
\eeq
In this case, we obtain:
\beq
I = 2 l^2 \, \int_0^{\infty} \mathrm{d} k \, \frac{ k^{(2l-1)} \Big[ 2 k^{2l} - (E + \bar{E}) \Big]}{(1+ (k^{2l} -E)(k^{2l} -\bar{E}))^2} \, .
\label{res2}
\eeq
This integral yields $I = \frac{l}{|E|^2 + 1}$, a pretty unexpected result, since in general a integer winding number should be expected. 
However this result can be explained quite naturally by analyzing 
 the map (\ref{psimod}) directly. This map can be expressed as: 
\beq
\omega_{\mathbf{k}} = \frac{1}{k^{2l}- E} \, e^{i \phi_k l} \, .
\label{map2}
\eeq
As for (\ref{psi}), we can write again:
\beq
\omega_{\mathbf{k}} = \big(f_- (k) + f_+ (k) \big) \, e^{i \phi_k l} \, 
\eeq
with $ f_- (k) =  \frac{1}{k^{2l}- E }$, $k < E^{1/2l}$, and 
$ f_+ (k) =  \frac{1}{k^{2l}- E}  $, $k > E^{1/2l}$.
We notice that $f_- (k) \, e^{i \phi_k l}$ is homotopic to a constant map $\tilde{f}_- (k) = c$,
since $f_- (k) = (-\infty, -\frac{1}{E}]$ (the minus sign is re-absorbable in the phase $\phi_k$ ) and not every point of the target stereographic plane $R^2$ is 
covered by $f_- (k) \, e^{i \phi_k l}$. 
Then we can write 
\beq
\omega_{\mathbf{k}} = \big(f_- (k) + f_+ (k) \big) \, e^{i \phi_k l}  \sim  f_- (k)  \, e^{i \phi_k l} 
\eeq
(here the symbol $\sim$ means "continuously deformable to"). 
Since again $f_+ (k) = [0, \infty)$ and is monotonic, it yields a contribution $I_+ = l$ to $I$ for every value of $E$.
However $ f_- (k)$ gives a non vanishing contribution to $I$, covering a part of the sphere with area
\beq
I_- = - \frac{1}{\pi} \, \int_{0}^{\frac{1}{E}}  \mathrm{d}k  \, \frac{2 \pi \, k}{(1+k^2)^2} = - \frac{E^2}{E^2 +1} \, ,
\eeq
where the minus sign appears since $|f_-(k\to \infty)| \to \infty$.
This contribution sums up to $I_+$, giving the result (\ref{res2}):
\beq
I = I_+ + I_- = l - l \, \frac{E^2}{E^2 +1} = l \,   \frac{E^2}{E^2 +1} \, .
\eeq
In spite of the value of $I$, the real winding number related to (\ref{psimod}) is $\tilde{I} = I_+ = l$, since we know  that $f_- (k)$ is homotopic to a constant map, 
a fact also resulting in a value of $|I_-|$ smaller than 1.\\
This result matches  the fact that the BCS case and the (\ref{map2}) case are linked by the transformation in the gap 
$\Delta  \to \Delta (k) = \Delta \, e^{i \phi_k l} $. However, this map is continuous but not invertible, wrapping $l$ times: this is the
reason of $\tilde{I} = l$. 

In conclusion, the case (\ref{map2})  describes a phase with winding number $I = l \frac{N}{2}$ (with $\frac{N}{2}$ being the number of Cooper pairs in the ground-state). However, the energy of Bogoliubov quasiparticles is the same as in the BCS case, and always gapped; thus no  phase transitions arise
and the system is always in a phase with nontrivial topology.


\begin{thebibliography}{xx}
\bibitem{notevan}
This is also suggested by the fact that $\mathrm{lim}_{k \to \infty} \, \frac{2 k^{2n} -2 \mu + \Delta^2\, k^{2(l-n)}}{2 \, \lambda_k}  = o(1)$ only if $l < 2 n$ (see Eq. \eqref{ffin}).
\bibitem{noteint}
If a lattice realization for Eq. \eqref{Ham} is considered, the same integral must be performed on the corresponding
Brillouin zone.
\end{thebibliography}
\end{document}